\documentstyle[11pt,paspconf,epsfig]{article}

\newcommand{\uband}{$U$}
\newcommand{\bband}{$B$}
\newcommand{\vband}{$V$}
\newcommand{\rband}{$R$}
\newcommand{\gband}{$g$}
\newcommand{\iband}{$I$}

\newcommand{\hband}{$H_\alpha$ 656.5nm}

\newcommand{\steidU}{\ifmmode{{U_n}}\else{${U_n}$}\fi}
\newcommand{\steidG}{\ifmmode{{G}}\else{${G}$}\fi}
\newcommand{\steidR}{\ifmmode{\mathcal{R}}\else{$\mathcal{R}$}\fi}

\newcommand{\gtabout}{\stackrel{>}{_{\sim}}}

\newcommand{\sixc}{6C\,0140$+$326}
\newcommand{\eightc}{8C\,1435$+$635}
\newcommand{\btwo}{B2\,0902$+$343}
\newcommand{\fourc}{4C\,41.17}

\newcommand{\lya}{Ly$\alpha$}
\newcommand{\Ha}{H$\alpha$}

\newcommand{\focas}{\textsc{focas}}

\begin{document}

\title{Searches for galaxies at $z \gtabout 4$ through Lyman--limit imaging}

\author{R.\ Stevens, M.\ Lacy and S.\ Rawlings}%
\affil{Astrophysics, Department of Physics, Keble Road, Oxford, 
OX1 3RH}

\begin{abstract}
We present preliminary results of a search for galaxies at $z
\gtabout 4$ through Lyman--limit imaging of the fields of known
high--redshift radio--galaxies.  Objects were selected by means of
their broad--band colours, and spectroscopy of candidate objects in
one of the fields has been performed through multi--slit spectroscopy
at the 4.2m William Herschel Telescope.  These spectra show some of
the first $z>4$ galaxies to be identified using the Lyman break
technique.
\end{abstract}


\section{Introduction}

Encouraged by the early results of Steidel and others (e.g.
Steidel \& Hamilton, 1993), and of our own pilot programme imaging a 1.5
arcmin$^2$ region around the radio galaxy \fourc\ (Lacy \& Rawlings,
1996), we embarked in November 1995 upon a project to apply the
technique of Lyman--limit imaging to fields of area $\sim$40
arcmin$^2$ around high--redshift radio--galaxies using the Prime
Focus of the WHT in La Palma.  

Our initial aim was to identify galaxies at redshifts similar to
those of the central radio--galaxy in each field, searching for $U$
dropouts in the $3 <z <4$ fields and $B$ dropouts in those at $z>4$,
with spectroscopic follow-up where possible.  We imaged four
radio--galaxy fields in total, two of the radio galaxies being at
$z>4$.  Our imaging observations are summarised in
Table~\ref{tab:obs}.

\begin{table}
\begin{center}
\caption{Imaging observations}
\label{tab:obs}
\scriptsize
\begin{tabular}{lrrrrrc}
\\
\tableline
Field & Redshift &\multicolumn{1}{c}{Filter} & \multicolumn{1}{c}{Date} & 
\multicolumn{1}{c}{Seeing} & \multicolumn{1}{r}{Total integration} 
& Sensitivity \\
& & & & \multicolumn{1}{c}{(arcsec)} & \multicolumn{1}{r}{time (s)} & limit\tablenotemark{a}\\
\tableline \tableline

\btwo   & 3.40 & \uband & 20--21.11.95     & 1.3--2.3 & 18000 & 26.3 \\
\multicolumn{2}{l}{(Lilly, 1988)} 	& \bband & 21.11.95     & $\approx$2.2 & 1600 & 25.9 \\
 &        & \rband & 20.11.95     & $\approx$1.3 & 2000 & 26.0 \\
\tableline
\fourc & 3.80 & \uband & 22.11.95     & $\approx$1.7 & 18000 & 26.3 \\
\multicolumn{2}{l}{(Chambers  {\it et~al.}, 1990)} & \gband & 22.11.95     & 1.9--2.3 & 2800 & 26.5
 \\
&        & \rband & 21.11.95     & 1.4--1.8 & 2400 & 26.0 \\
\tableline
\sixc & 4.41 & \bband & 21--22.11.95     & 1.2--1.8 & 12600 & 27.2 \\
\multicolumn{2}{l}{(Rawlings {\it et~al.}, 1996)}
& \vband & 20.11.95  & $\approx$1.3 & 1800 & 25.7\\
&        & \rband & 21.11.95  & $\approx$1.4 & 1800 & 26.1\\
&        & \iband & 20.11.95  & 1.1--1.4 & 6600 & 25.7 \\
&        & \Ha{\tablenotemark{b}} & 21.11.95  & $\approx$1.2 & 7200 & 24.1
 \\

\tableline
\eightc & 4.25 & \bband & 19.04.96     & $\approx$1.0 & 5400 & 27.0 \\
\multicolumn{2}{l}{(Lacy {\it et~al.}, 1994)} & \rband & 19.04.96     & $\approx$1.1 & 2700 &
26.2 \\
&        & \iband & 19.04.96     & 0.9--1.1 & 5400 & 25.8 \\
\tableline

\end{tabular}

\end{center}

\tablenotetext{a} { $2.5\sigma$ limit in AB magnitudes for
1.5\arcsec\ radius aperture.  In practice galaxies will be detected below
these limits if they are sufficiently compact for significant excess flux
to be recorded in at least six contiguous pixels.
}
\tablenotetext{b}{ This narrow-band filter was a redshifted
\Ha\ filter with peak response at 656.5nm.
This wavelength corresponds approximately to the wavelength of the
\lya\ line at the redshift of \sixc.}

\end{table}

\section{Selection of candidate $z\gtabout4$ galaxies}

\begin{figure}

\begin{minipage}[b]{1\linewidth}
\begin{minipage}[b]{.49\linewidth}
\centering\epsfig{file=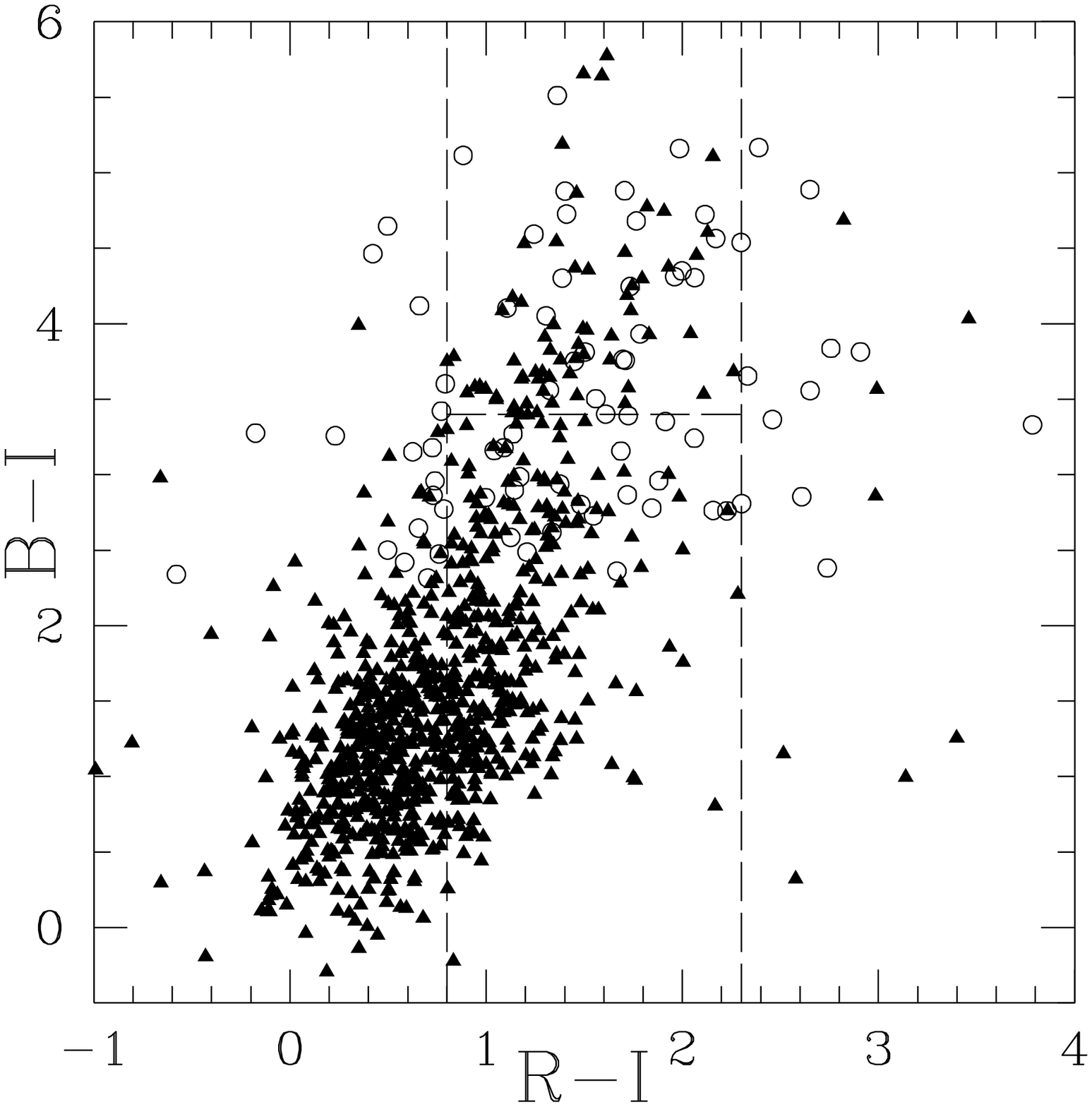, width=65mm, height=65mm}
\end{minipage}
\hfill
\begin{minipage}[b]{.49\linewidth}
\centering\epsfig{file=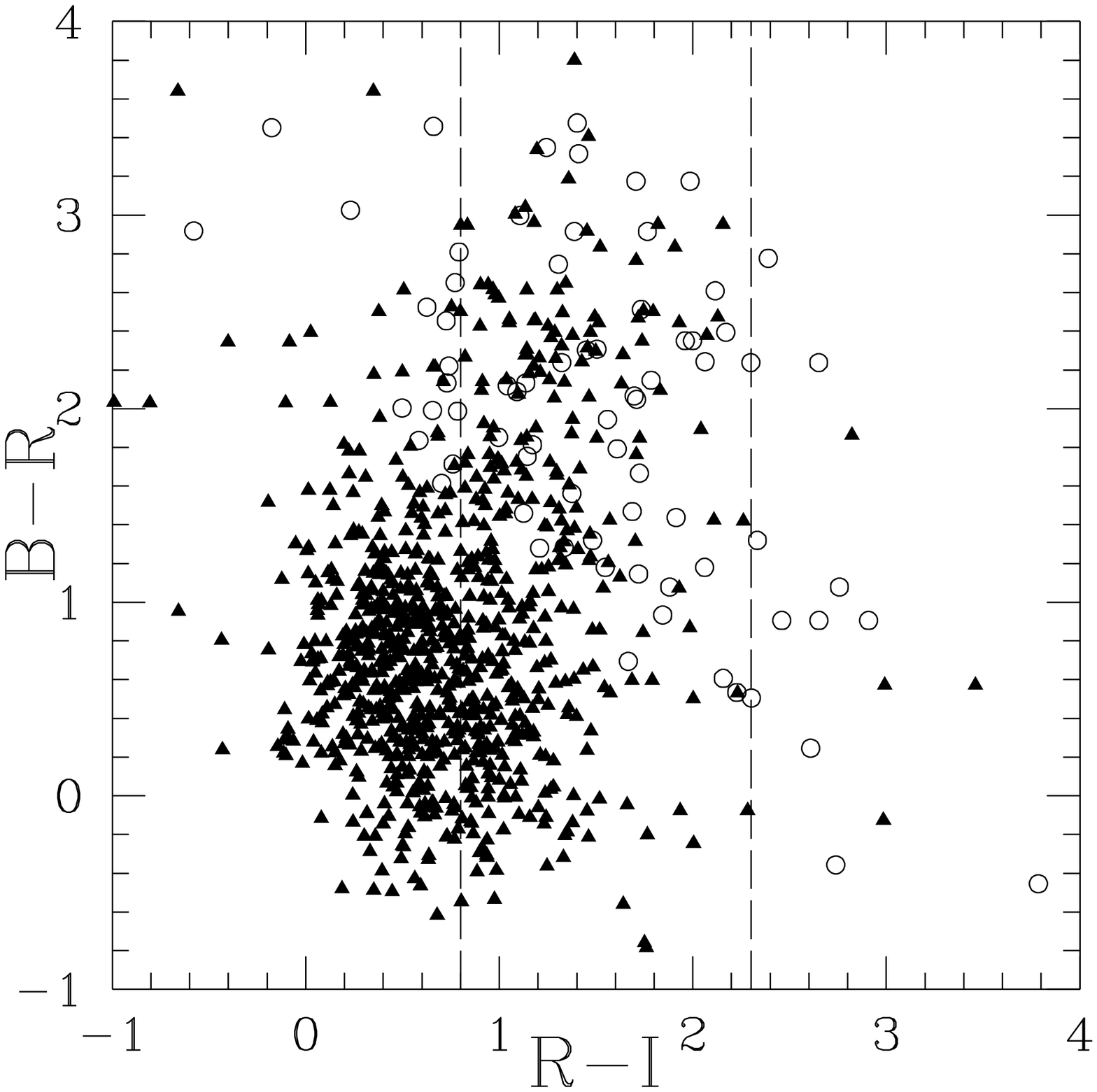, width=65mm, height=65mm}
\end{minipage}
\end{minipage}

        \caption{Colour--colour plots for faint objects ($21.0 < I < 24.5$)
in the field of \sixc.
Triangles are objects which have been detected by \focas\ (Jarvis
\& Tyson, 1981) in
\bband, \rband\ and \iband; circles show $B$ dropouts.  
The dashed lines show the limits of our original selection criteria (\bband\
dropouts below the $B-I$ limit are also included as candidates). 
Selection criteria will be reviewed following analysis of real and model
spectra, taking into account intergalactic absorption.
	}
        \label{fig:colcol}
\end{figure}

We based our initial selection criteria on the assumption that
high redshift galaxies are likely to show a relatively flat spectrum at
wavelengths longward of \lya, a pronounced spectral break at \lya, and
little or no flux below the Lyman limit.  For simplicity, we use the
AB magnitude system.

Initial candidates were selected according to the following criteria:
$21.0 < I <
24.5$ (to eliminate the very faintest objects); no detectable flux in $B$,
or $B-I > 3.4$ (with hindsight, a cutoff in $B-R$ would probably
have been better); and $0.8 < R-I < 2.3$, to eliminate objects which are
obviously merely very red objects.  There remains a possible contamination
from intrinsically red objects (e.g. M stars) and galaxies at lower
redshift, particularly those with a strong 4000\AA\ break.

\begin{figure}
\begin{center}
\begin{minipage}[t]{1\linewidth}
\begin{minipage}[t]{.25\linewidth}
\centering\epsfig{file=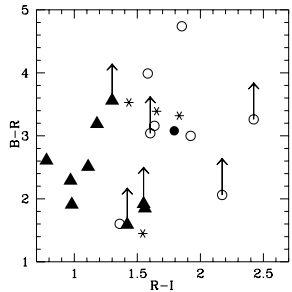, width=40mm, height=40mm}
\end{minipage}
\hfill
\begin{minipage}[b]{.74\linewidth}
        \caption{Colour-colour plots of the objects for which we have been
able to attempt identification from their spectra.  Open circles represent
objects which are most likely at $0.3 < z < 0.8$, asterisks are
almost certainly stars and the solid circle is possibly a QSO at $z=1.34$.
Triangles are objects for which a high--redshift nature, although not
necessarily confirmed, cannot be ruled out.
	}
\vspace{5mm}
        \label{fig:specbrri}
\end{minipage}
\end{minipage}
\end{center}
\end{figure}

Colour--colour plots for the magnitude ranges in question are shown in
Figure~\ref{fig:colcol}. Those for objects for which spectra are
available are shown in Figure~\ref{fig:specbrri}.
In the light of the results from our spectroscopic observations and from
detailed modelling of the {\em expected} colours of $z \gtabout 4$
galaxies, we are looking to see whether our selection criteria can be
improved --- there is a strong suggestion that we are selecting objects too
red in $R-I$ (Figure~\ref{fig:specbrri}).

\section{Spectroscopy in the field of \sixc}

\begin{figure}
\begin{center}
\centering\epsfig{file=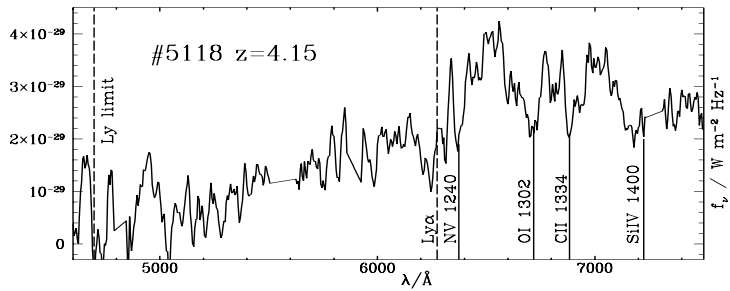, width=10cm, height=4cm}
        \caption{13645s WHT spectrum of one of our candidates, 
smoothed with a 25\AA\
boxcar.  Some relics of the sky subtraction have been clipped. 
Although noisy, we believe this spectrum to show a galaxy at a redshift
of $4.15$, and thus one of
the first $z>4$ galaxies to be found using the Lyman--limit technique.
	}
        \label{fig:specs}
\end{center}
\end{figure}

Using the LDSS2 multi-slit spectrometer at the WHT, we
took spectra of 49 objects in the \sixc\ field, 42 selected
by means of their broadband colours and seven through detection of excess
flux through our \hband\ filter with respect to the broad \rband.  We
split the objects across two slit masks according to their
\rband\ magnitudes, with total integration times 13645s for the brighter
objects and 26000s for the fainter ones.

Of these 49, we were unable to determine any significant features on 25 of
the spectra, either due to insufficient integration time or through
problems with the slit mask such as ghosting or stray light from the
alignment stars.  Of the remaining 24 for which we were able to deduce
anything of the nature of the object, we find that six are stars, eight
are almost certainly galaxies at relatively low redshift ($0.3 < z < 0.8$)
and one is most likely a QSO at $z=1.34$.

Two have spectra with prominent breaks and absorption features consistent with
their being $z > 4$ galaxies, one is presented in
Figure~\ref{fig:specs}.  We believe that these are among the first to be
found using the Lyman--limit technique at $z > 4$.

The remaining seven spectra have prominent continuum breaks and/or
spectral features not inconsistent with their being at high redshift,
although in many cases the possibility that they are low-redshift
objects cannot be ruled out.

\begin{table}
\begin{center}
\caption{Possible high--redshift objects as determined from spectra}
\label{tab:spec}
\scriptsize
\begin{tabular}{rrrrrrc}
\\
\tableline
I.D. &	\multicolumn{1}{c}{$B$} &	\multicolumn{1}{c}{$V$} &	\multicolumn{1}{c}{$R$}	& \multicolumn{1}{c}{$I$}	& \multicolumn{1}{c}{6565\AA}&	Possible $z$ \\
\tableline
\tableline
1028 &	26.44 &	24.64 &	23.93 &	22.82 &	23.17 &	0.58 or 4.09 \\

& \multicolumn{6}{l}{
$K=$22.8, hence $I-K = 0.0 \pm 0.25$ and very flat above continuum break }\\
122 &	25.55 &	25.17 &	23.70 &	22.14 &	24.08 &	4.02? \\
& \multicolumn{6}{l}{
Based on possible identification 
of 4 absorption features:  
N{\sc v} 1240, C{\sc i} 1277,  O{\sc i} 1302, } \\ 
& \multicolumn{6}{l}{
C{\sc ii} 1334 } \\
5099 &	$>$27.17 &	24.74 &	23.61 &	22.31 &	$>$24.1 & 0.64 or 4.4 \\
& \multicolumn{6}{l}{
From continuum break only} \\
5118 &	25.96 &	24.19 &	23.67 &	22.70 &	23.40 &	4.15 \\
& \multicolumn{6}{l}{
From N{\sc v} 1240, O{\sc i} 1302, C{\sc ii} 1334, Si{\sc iv} 1400} \\

5123 &	25.06 &	24.06 &	23.15 &	22.17 &	22.47 &	0.58 or 4.1 \\
63 &	25.98 &	24.30 &	23.37 &	22.59 &	23.02 &	3.34?? \\
84 &	26.86 &	25.59 &	23.67 &	22.49 &	$>$24.1 & 0.6 or 4.3 \\
127 &	$>$27.17 &	$>$26.1 &	25.58 &	24.16 &	$>$24.1 & 0.55 or 4.1 \\
95 &	$>$27.17 &	$>$26.1 &	25.25 &	23.70 &	$>$24.1 & 4.58?? \\
& \multicolumn{6}{l}{
No continuum detected in spectrum: strong 
emission at 6785\AA\ (possibly \lya?)
}  \\
\tableline

\end{tabular}
\end{center}
\end{table}

\section{Acknowledgements}

Our thanks go to Richard McMahon for the loan of his \gband\ band filter.
The WHT is operated on the island of La Palma by the Royal Greenwich
Observatory in the Spanish Observatorio del Roque de los Muchachos of the
Instituto de Astrof\'{\i}sica de Canarias.  REJS acknowledges the support
of a PPARC studentship.



\begin{references}

\reference
Chambers K.~C.  {\it et~al.}, 1990.
\newblock ApJ, 363, 21.

\reference
Jarvis J.~F. and Tyson J.~A., 1981.
\newblock AJ, 86, 476.

\reference
Lacy M.~D., Rawlings S., 1996.
\newblock MNRAS, 280, 888.

\reference
Lacy M.~D. {et~al.}, 1994.
\newblock MNRAS, 271, 504.

\reference
Lilly S.~J., 1988.
\newblock AJ, 333, 161.

\reference
Rawlings S. {et~al.}, 1996.
\newblock {\it Nature}, 383, 502.

\reference
Steidel C.~C., Hamilton D., 1993.
\newblock AJ, 105, 2017.

\reference
Steidel C.~C. {et~al.}, 1996b.
\newblock ApJL, 462, L17.

\end{references}
\end{document}